\documentclass{statsoc}

\usepackage[a4paper]{geometry}
\usepackage{graphicx}
\usepackage[textwidth=8em,textsize=small]{todonotes}
\usepackage{amsmath}
\usepackage{natbib}
\usepackage{booktabs} 
\usepackage{algorithm} 

\newtheorem{theorem}{Theorem}
\newtheorem{lemma}[theorem]{Lemma}
\newtheorem{corollary}{Corollary}[theorem]
\newtheorem{definition}{Definition}
\newtheorem{proposition}{Proposition}

\usepackage{amssymb}
\usepackage{algorithmicx}
\usepackage[noend]{algpseudocode}

\title[In-Betweenness]{A Measurement of In-Betweenness and Inference Based on Shape Theories}

\author[Dustin Pluta {\it et al.}]{Dustin Pluta}
\address{Department of Statistics, University of California,
Irvine,
USA.}
\author{Xiangmin Xu}
\address{Department of Anatomy and Neurobiology, Department of Biomedical Engineering, Department of Computer Science, The Center for the Neurobiology of Learning and Memory, The Center for Neural Circuit Mapping, University of California, Irvine, USA.}

\author{Daniel L. Gillen}
\address{Department of Statistics, University of California,
Irvine,
USA.}

\author{Zhaoxia Yu}
\address{Department of Statistics, University of California,
Irvine,
USA.}
\email{zhaoxia@ics.uci.edu}

\begin{document}
\begin{abstract}
We propose a statistical framework to investigate whether a given subpopulation lies between two other subpopulations in a multivariate feature space. This methodology is motivated by a biological question from a collaborator: Is a newly discovered cell type between two known types in several given features? We propose two in-betweenness indices (IBI) to quantify the in-betweenness exhibited by a random triangle formed by the summary statistics of the three subpopulations. Statistical inference methods are provided for triangle shape and IBI metrics. The application of our methods is demonstrated in three examples: the classic Iris data set, a study of risk of relapse across three breast cancer subtypes, and the motivating neuronal cell data with measured electrophysiological features. 
\end{abstract}

\section{Introduction}
Betweenness and similar measures are important concepts in network analysis \citep{borgatti2009network}. For example, the closeness centrality of a node quantifies the degree of closeness of the node to other nodes by summing up the lengths of shortest path, also known as geodesic, from the node to each of all other nodes \citep{bavelas1950communication}. It can be considered as a measure of broadcasters. Betweenness centrality is a related but distinct measure. It aims to find potential ``bridges'' rather than ``broadcasters''. Mathematically, the betweenness centrality of a node is defined as the number of times that it is on the shortest path between other pairs of nodes \citep{freeman1977set, bavelas1948mathematical}. These metrics have been widely used to understand the roles of individual vertices in a social network by examining the positions of vertices in a graphical network model \citep{borgatti2009network}.

Betweenness is also of high relevance in comparing multiple items or populations. A motivating example for the methodology developed here is a set of electrophysiological measurements collected from two neuronal populations with distinct functional and physiological characteristics, and a novel population believed to have a functional role overlapping with both of the existing populations. The two existing populations are parvalbumin(PV) expressing neurons, which tend to be fast-spiking, and cholecystokinin(CCK) expressing neurons that are non-fast-spiking. Recently, our collaborator Dr. Xu at the department of Neurology of UCI and his team found that PV/CCK double positive cells exist in adult mice. An important question is whether the PV/CCK neuronal population ``lies between'' PV and CCK populations with respect to a set of electrophysiological characteristics. 

Another interesting example is the iris data, which is a classical multivariate data set that has been widely used to illustrate various statistical analysis and machine learning, such as clustering and classification, of multivariate data. Fisher introduced the data to illustrate linear discriminant analysis \citep{fisher1936use}. It is perhaps less well known that the data were collected by Edgar Anderson to quantify the morphological features of three iris species: Iris setosa, Iris versicolor, and Iris virginica \citep{anderson1936species}. Based on four morphological features, namely petal length, petal width, sepal length, and sepal width, Anderson hypothesized that \textit{Iris versicolor} is ``in an intermediate position morphologically'' between \textit{Iris virginica} and \textit{Iris setosa}. In this case, the geometric relationship of interest falls into the general idea of in-betweenness.

In both examples, multiple features were measured on each subject. Although various dimension reduction methods have been used to visualize the geometric relationship between different subtypes of data, there is a lack of formal definition, quantification, and statistical framework to make inference of in-betweenness with respect to a set of observed features. We introduce a general method for statistical inference of in-betweenness, and consider two statistics motivated by random triangle theory.  We also consider the construction of bootstrap confidence regions for shape space parameters, which are of use when the relative positioning of three subgroup centroids is of interest.

This rest of this chapter is outlined as follows. In Section 2 we introduce the notations and several coordinates for random triangles and derive the distribution of random triangles formed from iid $N(0,1)$ observations. Metrics for quantifying in-betweenness and their statistical inference are provided in Section 3. The statistical approaches to make inference of in-betweenness are illustrated using simulations and three real examples in Section 4. We conclude this chapter with a discussion of the advantages, limitations, and future work in Section 5.

\section{Statistical Shape Theory of Random Triangles}
Interests in random triangles date back to at least 1884 with the publication of Lewis Carroll's ``Pillow Problem" \citep{carroll1893curiosa}. Paraphrased, this problem asks: what is the probability that a randomly generated triangle in the plane is obtuse?  Despite its simplicity (and ambiguity), this question exemplifies the perspective of statistical shape theory, which is concerned with the stochastic properties of random configurations of points (landmarks) when location, scale, and orientation have been removed.

A modern development of statistical shape theory was initially motivated by applications to studying the relationships of shape and size in biological specimens.  In this setting, the goal is to conduct inference regarding the geometric characteristics of biological features, such as skull shape across samples of closely related species.  For this analysis, relevant landmarks are labeled on each biological specimen, yielding a sample of observed shapes.  Given such a sample, the tools of statistical shape theory can be used to test equality of shapes across species, or quantify the morphological similarity of different subpopulations.  Other examples of previous applications of shape analysis include the study of vertebrae from a sample of specimens from the same species \citep{mardia1989shape}, protein molecules \citep{green2006bayesian}, and magnetic resonance images \citep{dequardo1996spatial}.  In these settings, the shapes under consideration are two- or three-dimensional, with the landmarks chosen to sufficiently describe the physical features of scientific interest.

Different from the usual application of shape theory to a sample of observed, physical objects, the present work instead applies the results of classical shape theory to analyze the relationship of three subpopulations measured across a set of common variables.  This approach is similar in some ways to correlation analysis of two feature sets, where the joint relationship is quantified by a scale-, location-, and rotation-free triangle, rather than a correlation coefficient.

\subsection{Triangle Shape Space}
In this section, we review the relevant definitions for general shape theory and some of the existing results regarding triangle shape space.  We mostly follow the terminology and definitions established in \cite{dryden2016statistical}, which provides a comprehensive introduction to statistical shape theory.

\begin{definition}
A \textbf{configuration} is a set of $k$ points (landmarks) in $\mathbb{R}^p$. The \textbf{configuration matrix} $X$ is the $k \times p$ matrix of the landmark coordinates.  The \textit{configuration space} is the space of all configuration matrices.
\end{definition}

To construct a formal definition of \textit{shape}, we first consider the \textit{pre-shape} of a configuration, which is the remaining information after location and scale have been removed. 
\begin{definition}
For a $k \times p$ configuration matrix $X$, the \textbf{pre-shape} $Z$ is
\begin{equation}
Z = \frac{H X}{\|H X\|}
\label{eqn:preshape}
\end{equation}

where $\|H X\| = \sqrt{\text{tr}(X'H'H X)}$, for $H$ the $(k - 1) \times k$ Helmert submatrix.

The \textbf{pre-shape} space of $k$ landmarks in $p$ dimensions, denoted $S^k_p$, is the set of all possible pre-shapes $Z$ over configurations $X \in \mathbb{R}^{k \times p}$.
\label{def:pre-shape}
\end{definition}

\textbf{\textit{Remark}} In the case of triangular configurations, we denote the $2 \times 3$ Helmert submatrix $\Delta$, that has entries 
\begin{equation}
\label{eqn:helmert}
\Delta = \begin{pmatrix}\frac{1}{\sqrt{2}} & \frac{-1}{\sqrt{2}} & 0 \\
\frac{1}{\sqrt{6}} & \frac{1}{\sqrt{6}} & \frac{-2}{\sqrt{6}}\end{pmatrix}.
\end{equation}

$\Delta$ can be viewed as the edge matrix of an equilateral triangle, and plays an important role in the construction of triangle shape coordinates given in Section \ref{shape_coordinates}.

The \textit{shape} of a configuration is formally defined as the equivalence class of the pre-shape over all possible rotations.
\begin{definition}
The \textbf{shape} of configuration $X$ with pre-shape $Z$ is the equivalence class
\begin{equation}
    [X] = \{Z\Gamma ~|~ \Gamma \in SO(p)\},
\end{equation}
where $SO(p)$ is the group of $p\times p$ orthogonal matrices with positive unit determinant.
\end{definition}
\begin{definition}
The \textbf{shape space} of $k \times p$ configurations, denoted $\Sigma_k^p$ is the set of all equivalence classes $[X]$ for configurations $X \in \mathbb{R}^{k \times p}$.
\end{definition}
\begin{theorem}[\cite{dryden2016statistical}]\label{thm:triangle_shape_space}
For the case of triangular configurations ($k = 3, p \geq 2$), the pre-shape space is a hypersphere embedded in $\mathbb{R}^{2p}$.  Triangular shape space can be identified with the unit disk in $\mathbb{R}^2$, or, equivalently, with the upper hemisphere of radius 1/2 in $\mathbb{R}^3$.
\end{theorem}

A detailed discussion and derivation of the properties of triangle shape and pre-shape spaces is given in \cite{dryden2016statistical}.  The triangle shape coordinates in Section \ref{shape_coordinates} provide explicit mappings from a triangular configuration $X$ to triangle shape space.

\subsection{Triangle Shape Coordinates}
\label{shape_coordinates}
Many formulations of triangle shape coordinates are possible, such as those developed by  \cite{bookstein1986size},  \cite{kendall1984shape}, and \cite{dryden1991general}.  We adopt a set of polar shape coordinates based on the formulation of Kendall's spherical coordinates presented in \cite{edelman2015random}, which are the natural result of a specific mapping from configuration space to shape space, and which have a linear relationship with the squared triangle side lengths (after scaling).  The relationship of these coordinates to other shape coordinate systems is discussed in Section ~\ref{shape_coordinates}, with further details given in \cite{dryden2016statistical}.

To define triangle polar coordinates, we consider a transformation from a configuration $X$ to coordinates $(r, \phi), 0 \leq r \leq 1, 0 \leq \phi < 2\pi$ constructed by successively removing the location, orientation, and scale information.  To remove location information we can simply center the columns of $X$, and hereafter assume that $X$ has been centered.  To remove the scale and orientation, it is convenient to instead work with the \textit{edge matrix}, $E$, which contains the edge vectors defined by the configuration $X$.  $E$ can be calculated from $X$ by $E = X'T$, where $T$ is the pairwise difference matrix 
\begin{equation}
    T = \begin{pmatrix}
    1 & -1 & 0\\
    0 & 1 & -1\\
    -1 & 0 & 1
    \end{pmatrix}.
\end{equation}
Let $M=E\Delta'$ where $\Delta$ is the Helmert matrix given in \ref{eqn:helmert}. To remove location and reflection, consider the singular value decomposition of $M$
\begin{align}
    M &= UD V' \\
    &= U\begin{pmatrix}d_1 & 0\\0 & d_2\end{pmatrix}\begin{pmatrix}\cos(\phi/2) & \sin(\phi/2)\\-\sin(\phi/2) & \cos(\phi/2)\end{pmatrix}\label{eqn:svd},
\end{align}

\noindent where we assume $d_1 \geq d_2 \geq 0$.  Discarding the left eigenvectors $U$ removes the rotation and reflection information from the configuration.  The polar shape coordinates are then obtained from the residual transformation $DV'$.
\begin{definition}
\label{def:polar}
\textbf{Polar shape coordinates} for a triangular configuration with decomposition in Equation \ref{eqn:svd} are $(r, \phi) \in [0, 1] \times [0, 2\pi)$, with $$r = \sqrt{1 - 4d_1^2d_2^2/(d_1^2 + d_2^2)^2}$$\\
The corresponding \textbf{rectangular shape coordinates} are $u = r\cos\phi, v = r\sin\phi$.
\end{definition}

This mapping verifies the result that triangle shape space is identifiable with the unit circle in $\mathbb{R}^2$, and is independent of the dimension of the ambient space of $X$.  By construction, the shape representation of a given triangle is invariant to translation, rotation, and scaling of the original triangle, thus the transformation $X \to E \to M \to (r, \phi)$ associates every triangular configuration with a point on the unit disk such that configurations with the same shape are mapped to the same shape space point.  The singular exception is the case when all landmarks are coincident, which does not have defined shape space coordinates.

\begin{figure}
    \centering
\includegraphics[width=\textwidth]{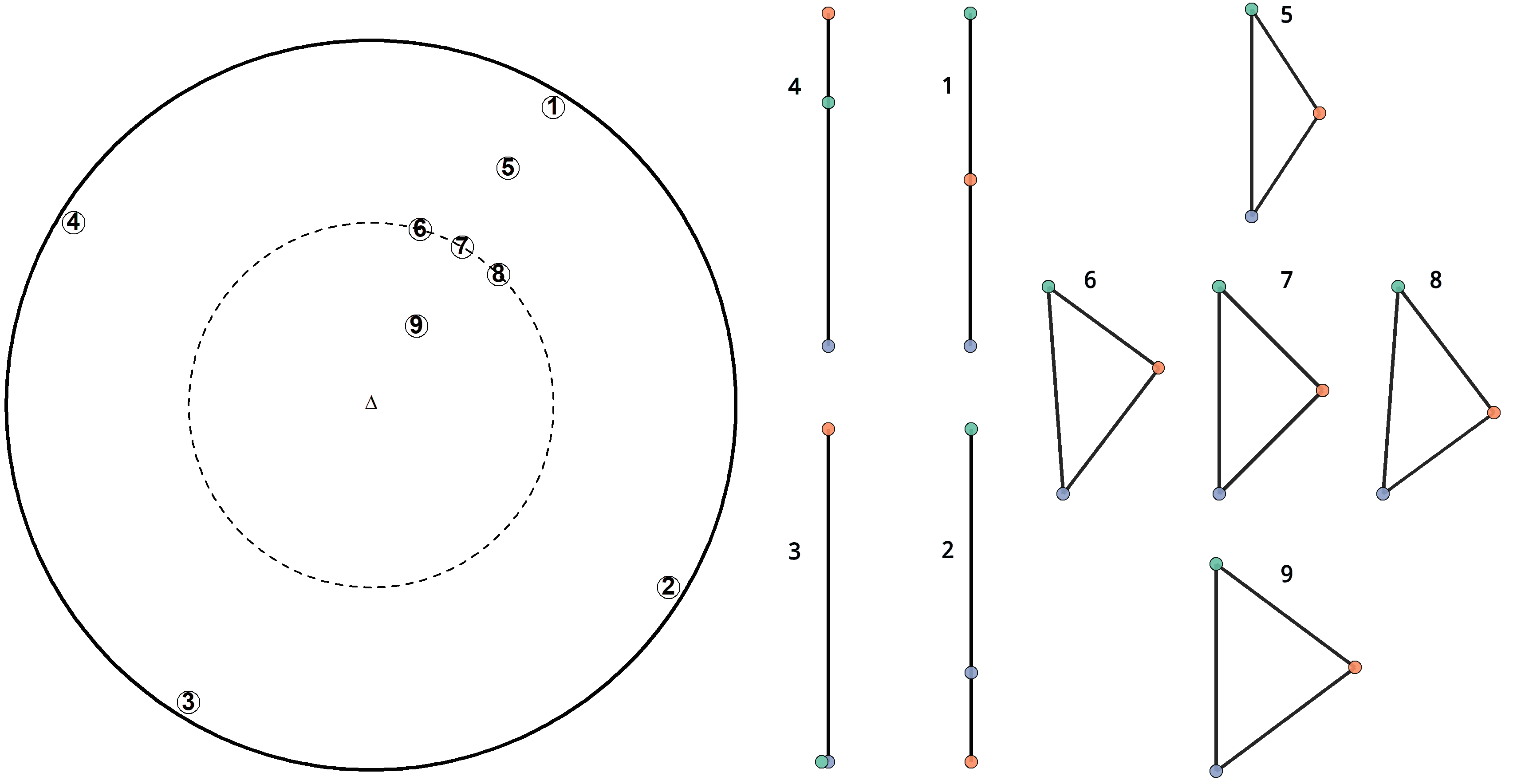}
    \caption[Example triangles and their shape space locations.]{(\textit{left}) A depiction of triangle shape space as the unit disc, with some example triangles labeled. The dashed line indicates the circle of radius $\frac{1}{2}$.  Triangles with equal shape space radius have equal standardized areas. (\textit{right}) The configurations corresponding to the labeled triangles.  The shape space boundary consists of the degenerate triangles. Shape space is continuous with respect to $r$ and $\phi$; small changes in coordinates correspond to small changes in triangle shape, as shown by the configurations for triangles 5 -- 9.}
    \label{fig:triangle_examples}
\end{figure}

Triangle shape space is structured with many intuitive properties.  Circles centered at the origin with radius $\leq 1$ describe triangles of equal area, with the boundary consisting of degenerate triangles of area 0, and the origin equal to the unique equilateral triangle. A radius in shape space (of points with equal angular measure $\phi$) represents different scalings of the same rotation of the equilateral triangle, which follows from the equality $d_1^2 = (r + 1) / 2$, after standardization.  Importantly, shape space is continuous with respect to shape, i.e., points close together in shape space represent approximately similar triangles.  \cite{edelman2015random} provides additional details on the structure of triangle shape space, and discusses the transformation from the configuration to shape space coordinates from multiple theoretical perspectives for the two-dimensional case.  Figure \ref{fig:triangle_examples} shows the locations of some example triangles in triangle shape space.

The shape space coordinates can also be viewed as a linear transformation of standardized edge lengths, which are defined as
\begin{align}
    a^2 &= \frac{\|X_B - X_C\|^2}{\|X_B - X_C\|^2 + \|X_A - X_C\|^2 + \|X_A - X_B\|^2}\\
    b^2 &= \frac{\|X_A - X_C\|^2}{\|X_B - X_C\|^2 + \|X_A - X_C\|^2 + \|X_A - X_B\|^2}\\
    c^2 &= \frac{\|X_A - X_B\|^2}{\|X_B - X_C\|^2 + \|X_A - X_C\|^2 + \|X_A - X_B\|^2},
\end{align}

\noindent for a configuration with landmarks $X_A, X_B, X_C$. From \cite{edelman2015random}, the standardized edge lengths are related to the shape space coordinates by
\begin{equation}
\label{eq:uv_ab}
    \begin{pmatrix}
        a^2 \\ b^2 \\ c^2 
    \end{pmatrix} = 
    \frac{1}{3}\left[
    \begin{pmatrix}
        \frac{1}{2} & \frac{-\sqrt{3}}{2}\\
        \frac{1}{2} & \frac{\sqrt{3}}{2}\\
        -1 & 0
    \end{pmatrix}
    \begin{pmatrix}
        u \\ v
    \end{pmatrix} + 
    \begin{pmatrix}
        1 \\ 1 \\ 1
    \end{pmatrix}
    \right].
\end{equation}

\subsubsection{Example of Triangle Shape Coordinate Transformation}

The following example using Fisher's Iris data illustrates the transformation from a configuration to shape space.  We consider the triangle formed by the mean sepal length and mean sepal width stratified by species.  The configuration matrix is
\begin{align}
    X &= \begin{pmatrix}5.01 & 5.94 & 6.59\\
        3.43 & 2.77 & 2.97
\end{pmatrix}
\end{align}

Multiplying the pairwise difference matrix $T$ produces the edge matrix 
\begin{equation}
    E = \begin{pmatrix}
    -1.582 & 0.930 & 0.652\\
    0.454 & -0.658 & 0.204.
    \end{pmatrix}
\end{equation}

The corresponding transformation matrix $M = E\Delta'$ is
\begin{align}M &= \begin{pmatrix}
0.915 & 0.319\\
0.081 & -0.233
\end{pmatrix} 
= U\begin{pmatrix}0.969 & 0 \\ 0 & 0.247\end{pmatrix}
\begin{pmatrix}\cos(0.107\pi) & -\sin(0.107\pi)\\
\sin(0.107\pi) & \cos(0.107\pi)\end{pmatrix}.\end{align}

\begin{figure}[h!]
    \centering
    \includegraphics[width=\textwidth]{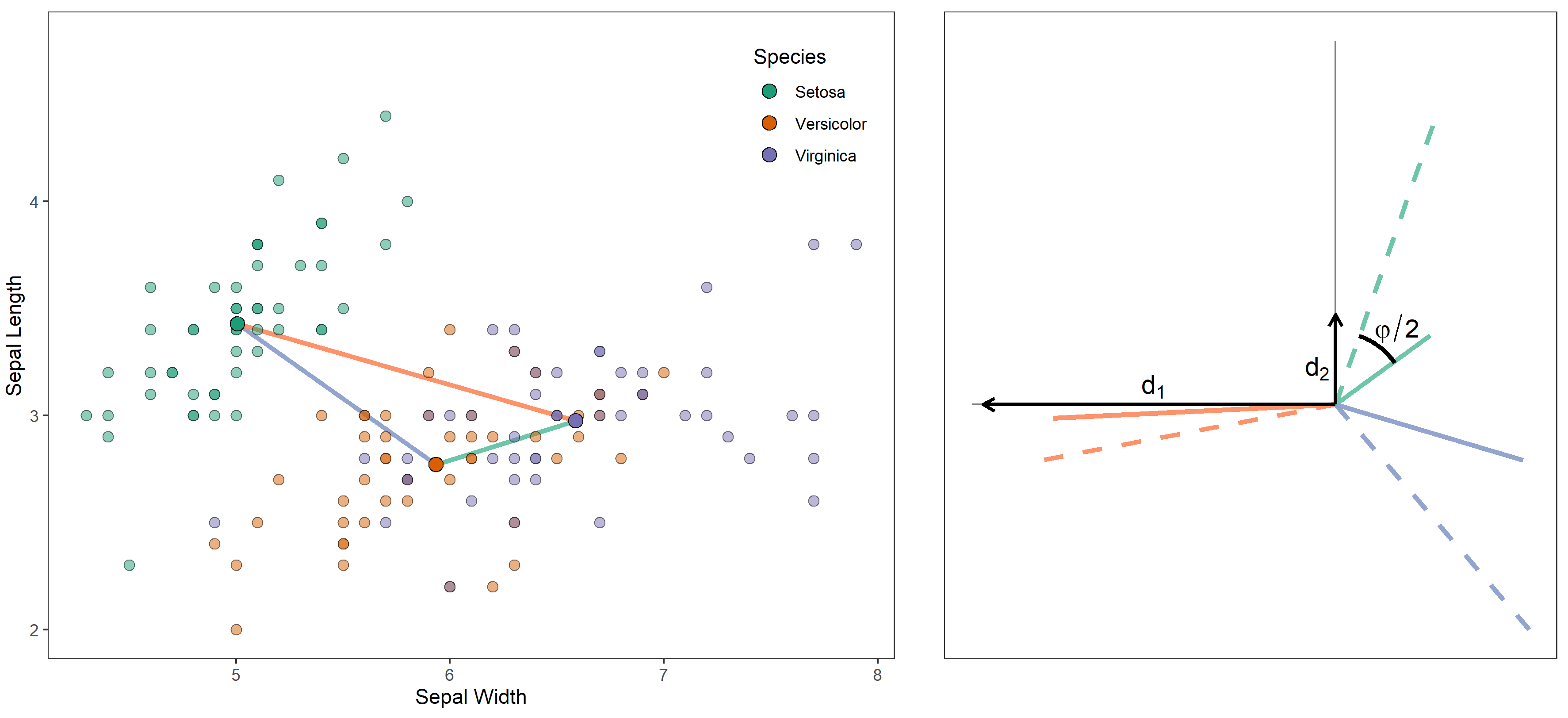}
    \caption[Iris example of mapping from a triangular configuration to shape space]{\textit{(left)} Observations from the Iris data set and the observed triangle formed by the group centroids for sepal width and sepal length. \textit{(right)} The observed triangle is mapped to polar shape coordinates by identifying the rotation ($\phi / 2$) and scaling ($d_1, d_2$) from the transformation matrix $M$.}
    \label{fig:triangle_transformation}
\end{figure}

Figure \ref{fig:triangle_transformation} shows the observed triangle and the identification of $\phi$ and $d_1, d_2$ to map the triangle to shape space.  The polar shape coordinates for this triangle are $(0.877, .214 \pi)$, representing a clockwise rotation of the equilateral edges by $0.107\pi$, and scaling along the new coordinates by $d_1 = 0.97, d_2 = 0.25$.

\subsection{Shape and Side Length Distributions}
\label{shape_and_side_distributions}

\noindent In this section, we present distributional results for triangles with landmarks generated by $X_A, X_B, X_C \overset{iid}\sim \mathcal{N}(\mu, \sigma^2I_p)$.  For the purposes of shape space analysis, this is equivalent to assuming the configuration matrix $X$ follows a standard $3 \times p$ matrix normal distribution, $X\sim \mathcal{N}(0, I_3, I_p)$.
\begin{lemma}
When the configuration distribution is $X \sim \mathcal{N}(0, I_3, I_p)$, the transformation matrix $M$ has distribution $\mathcal{N}(0, I_p, I_2)$.
\end{lemma}

\begin{lemma}
When the configuration distribution is $X \sim \mathcal{N}(0, I_3, I_p)$, the joint density of the polar shape coordinates $(r, \phi)$ is
\begin{equation}
    \label{eq:r_phi_pdf}
    f(r, \theta) = \frac{(p - 1)}{2\pi}r(1 - r^2)^{(p - 3)/2},
\end{equation}
with support $\{(r, \phi) | r \in [0, 1], \phi \in \mathbb{R}\}$.
\end{lemma}

\textit{Proof} Assume $X \sim \mathcal{N}(0, I_3, I_p)$, and let $d_1, d_2$ be the scaled eigenvalues of $M$. In this case, the ellipticity statistic has the form $2d_1d_2 = \sqrt{1 - r^2}$, with distribution function $P(\sqrt{1 - r^2} < x) = x^{p - 1}$ \citep{muirhead2009aspects}.  The pdf of $r$ can then be computed via variable transformation as 
\begin{equation}
\label{eq:r_pdf}
f_r(r) = (p - 1)r(1 - r^2)^{(p - 3)/2}.
\end{equation}

The distribution of $\phi$ can be deduced by observing that the distribution of $M$ is invariant under orthogonal transformations, thus the density $f_{\phi}$ must be constant; restricting the range gives $\phi \sim Unif(0, 2\pi)$.  The result then follows from Equation \ref{eq:r_pdf} and the independence of $r$ and $\phi$ in the spherical case \citep{muirhead2009aspects}.

\begin{theorem} 
When the landmark, the joint shape space distribution is
\begin{equation}
f_{u, v}(u, v) = \frac{(p - 1)}{2\pi}(1 - u^2 - v^2)^{(p - 3)/2},
\end{equation}
\end{theorem}
where $u$ and $v$ are defined in Definition \ref{def:polar}.

\noindent \textit{Proof} The distribution of $(u, v)$ induced by the iid normal configuration can be derived by computing the multivariate variable transformation of $(r, \phi)$ using the identities
\begin{align}
    r &= \sqrt{u^2 + v^2}\\
    \phi &= \arcsin\left(\frac{v}{\sqrt{u^2 + v^2}}\right).
\end{align}

The Jacobian of this transformation is
\begin{align}
    J &= \begin{vmatrix}
    \frac{u}{\sqrt{u^2 + v^2}} & \frac{v}{\sqrt{u^2 + v^2}}\\
    \frac{-v}{u^2 + v^2} & \frac{u}{u^2 + v^2}
    \end{vmatrix}\\
    &= (u^2 + v^2)^{-1/2}
\end{align}

This gives
\begin{align}
f_{u, v}(u, v) &= f_{r, \phi}(\sqrt{u^2 + v^2}, \arcsin\left(\frac{v}{\sqrt{u^2 + v^2}}\right))|J|\\ 
    &= \frac{1}{2\pi}(p - 1)\sqrt{u^2 + v^2}(1 - u^2 - v^2)\cdot \frac{1}{\sqrt{u^2 + v^2}}\\
    &= \frac{(p - 1)}{2\pi}\left(1 - u^2 - v^2\right)^{(p - 3)/2}.
\end{align}

\noindent The joint distribution of the squared side lengths follows from $f_{u, v}$ and the linear transformation relating $(a^2, b^2, c^2)$ and $(u, v)$ given by Equation \ref{eq:uv_ab}.
\begin{corollary}
When the configuration distribution is $X \sim \mathcal{N}(0, I_3, I_p)$, the joint squared side length distribution is

$$f_{a^2, b^2, c^2}(a^2, b^2, c^2) = \frac{3(p - 1)}{2\pi} \left(-\frac{1}{4} + a^2b^2 + a^2c^2 + b^2c^2\right)^{(p - 3)/2}.$$
\end{corollary}

The joint shape space distribution for $(u, v)$ is a form of square root Dirichlet distribution, i.e., the distribution of $(u^2, v^2)$ can be shown to follow a Dirichlet.
The marginal distribution of squared lengths can be derived using existing results about distributions of quadratics and their ratios \citep{gurland1953distribution}. However, for the joint distribution it is easier to obtain by transforming the joint distribution of $u$ and $v$.  These results are equivalent to existing results of shape distributions in other coordinate systems, but the explicit forms of these distributions for triangle polar coordinates and the scaled squared side lengths have not been previously presented to our knowledge.  

The above results can be considered as the standard distribution of triangle shapes, as the observations are iid from the standard normal distribution. For non-isotropic cases, such as those caused by non-standard variance-covariance of the features or different sample sizes, one can first standardize the observations. For example, suppose $x_{ij} \overset{iid} \sim \mathcal{N}(\mu, \Sigma)$ for $i = A, B, C, j = 1, \dots, n_i$. One can first subtract the mean $\bar x_{\dot\dot}$ and then standardize the variance-covariance by defining the transformed data: $\tilde x_{ij}=\Sigma^{-1/2} (x_{ij}-\mu)$. The triangle configuration will be defined using $X_i$'s,
where $X_i=\sqrt{n_i}\sum_{j=1}^{n_i}\tilde x_{ij}$ and all the distributional results hold asymptotically. The distribution with non-coincident landmark centroids will be deferred to a later section, as its distribution can be compactly expressed using Riemannian distance, which will be introduced in \ref{subsubsection:Riemannian}.

We have derived the distribution of the shapes of random triangles and expressed the distribution as functions of side lengths ($a^2$, $b^2$, $c^2$), unit disk polar coordinates $(r, \phi)$, or rectangular coordinates $(u,v)$. Although the distribution is theoretically important, the geometric characteristic of interest in our motivating examples is whether a particular group is in the middle of two other groups. In the following section, we propose metrics to quantify in-betweenness and study their statistical properties. 

\section{Quantifying In-betweenness and Shape Space Inference}

Recall that our purpose is to quantify in-betweenness and make statistical inference of it. We consider two measures of ``in-betweenness" for quantifying the hybridity of subpopulation B with respect to A and C: cosine of the supplementary angle corresponding to subpopulation B, denoted $\gamma$ and referred to as \textit{cosine in-betweenness}; and a shape space hybrid similarity statistic $\tau$, which is based on the intrinsic distance in the shape space. We refer to $\tau$ as the \textit{shape in-betweenness index (IBI)}.

\subsection{Cosine In-betweenness}

\begin{figure}[h!]
    \centering
    \includegraphics[width=0.8\textwidth]{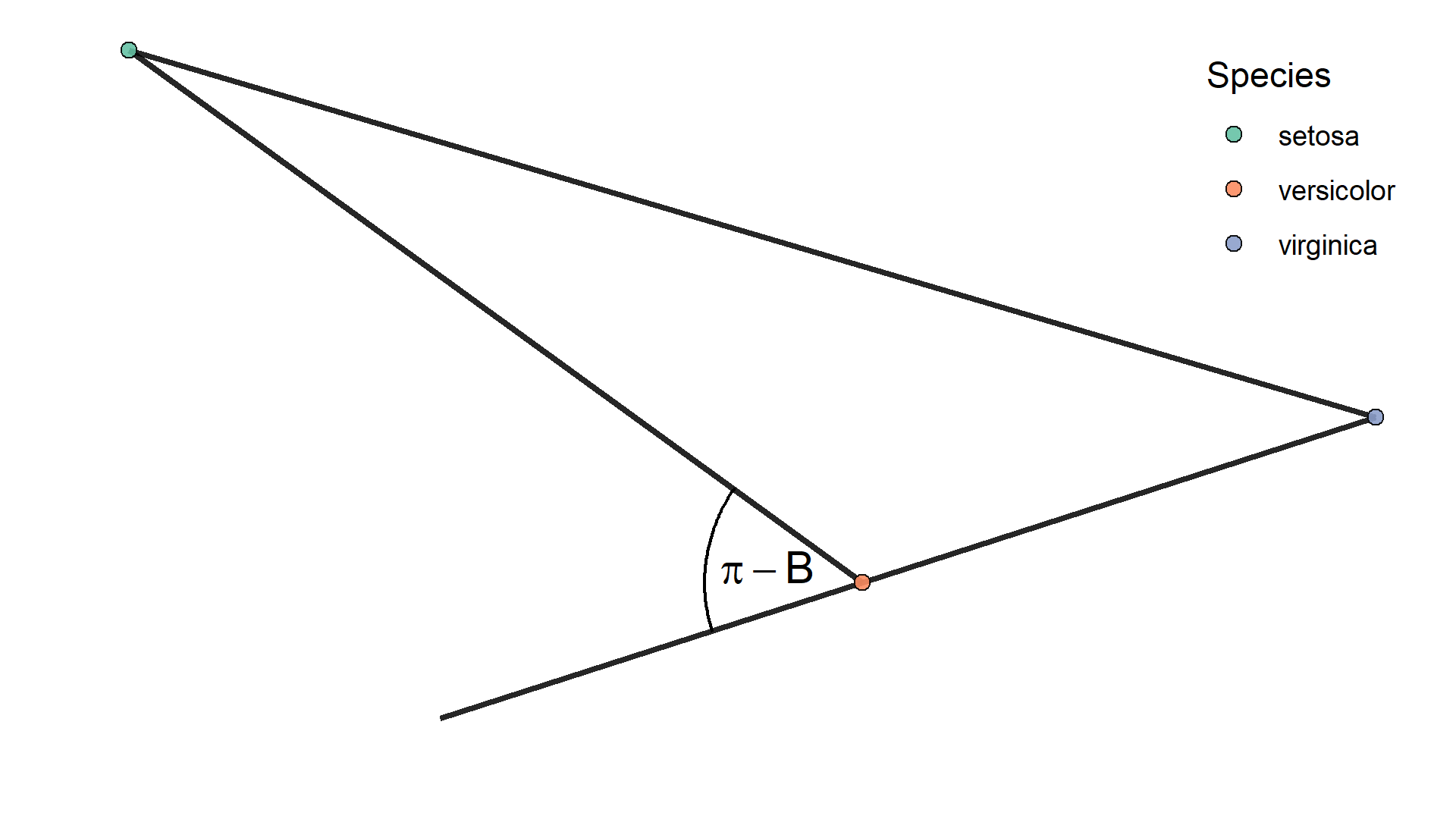}
    \caption[Definition of cosine in-betweenness]{Cosine in-betweenness to measure the hybridity of population 2 with respect to 1 and 3 is calculated as $\gamma = \cos(\pi - B)$.}
    \label{fig:iris_cosine_similarity}
\end{figure}

\noindent For a simple geometric approach to quantifying in-betweenness, we observe that degenerate triangles with $X_B$ in-between $X_A$ and $X_C$ have angle $B = \pi$, whereas degenerate triangles with $X_B$ not in-between $X_A$ and $X_C$ have $B = 0$.  This suggests the in-betweenness measure $\gamma = \cos(\pi - B)$, which yields $\gamma = 1$ for degenerate triangles with $X_B$ in-between $X_A$ and $X_C$, and $\gamma = -1$ for all other degenerate triangles. Triangles for which $B$ is close to $\pi$ are approximately degenerate with $X_B$ in-between $X_A$ and $X_C$, while triangles with $B$ close to 0 will be approximately degenerate with $X_B$ not in-between $X_A$ and $X_C$.  Cosine similarity is 0 for right triangles with $B = \pi / 2$.  The values of $\gamma$ over triangle shape space are shown in Figure \ref{fig:iris_cosine_similarity}.

\noindent The value of cosine in-betweenness can be computed in terms of the squared side lengths from the law of cosines:
\begin{align}
    \cos B &= \frac{a^2 + c^2 - b^2}{2ac}\\
    \gamma &= \cos(\pi - B) = \frac{2b^2 - 1}{2ac}\label{eq:gamma}
\end{align}

From this expression, we see that $\gamma$ has two discontinuities at $a = 0$ and $c = 0$, which are points on the disk boundary where $\gamma$ switches from -1 to 1.  Thus, although cosine in-betweenness is a simple and intuitive indicator of in-betweenness, it is not able to detect different degrees of in-betweenness, assigning values of 1 and -1 to triangles arbitrarily close together.

\subsection{Shape Space In-betweenness Index $\tau$}
\noindent To address the above issues with the cosine index, we instead propose an IBI that is continuous over shape space and sensitive to different degrees of in-betweenness. Again considering the in-betweenness of subpopulation B with respect to A and C, we motivate the definition by first assuming that the triangle with maximum in-betweenness should be the degenerate triangle with $B$ as the midpoint of $A$ and $C$ (or $b = 2a = 2c$), which we refer to as the midpoint triangle.  When the triangle sides are scaled so that $a^2 + b^2 + c^2 = 1$, the side lengths of the $B$-midpoint triangle are $a^2 = c^2 = \frac{1}{6}, b^2 = \frac{2}{3}$, with polar shape coordinates $(r, \phi) = (1, \pi/3)$ and Cartesian shape coordinates $(1/2, \sqrt{3}/2)$. We propose a shape space in-betweenness index defined as a transformation of the Riemannian shape distance between an observed triangle and the $B$-midpoint triangle.

\subsubsection{Riemannian Distance for Triangle Shape}
\label{subsubsection:Riemannian}

A useful notion of distance for triangle shapes can be defined via the Riemannian distance in the pre-shape space. From the geometric perspective, shapes are fibres on the pre-shape sphere, which inherit the pre-shape space Riemannian distance via projection of the fibres to points in shape space.  The formulation of pre-shape space given here is such that the projection is an isometric submersion of shape space in the pre-shape manifold, so that distances are preserved.  This distance is thus referred to as the \textit{Riemannian shape distance}, even though shape space is not a Riemannian manifold itself. Thus, the Riemannian distance for triangle shape is defined in terms of the \textit{pre-shape}, which is defined as the remaining information after location and scale have been removed from a configuration \citep{dryden2016statistical}.

Pre-shape space is a hypersphere in $\mathbb{R}^{(k - 1)p}$.  To see this, we observe from the definition of the pre-shape (Definition \ref{def:pre-shape}) that the coordinates of the pre-shape $Z = HX / \|HX\|$ are the standardized Helmertized coordinates of the configuration $X$, thus $Z$ has dimensions $(k - 1) \times p$ and satisfies $\|Z\| = 1$, and consequently pre-shape space is a sphere in $\mathbb{R}^{(k - 1)p}$.  We can therefore consider pre-shape space as a Riemannian manifold, and use its intrinsic Riemannian metric to induce a metric on shape space, which is a quotient space of the pre-shape sphere.

Geodesics on the pre-shape sphere are great circles, with the geodesic distance between pre-shapes $Z_1, Z_2$ defined as the shortest arc length along a great circle between $Z_1$ and $Z_2$ \citep{terras2013harmonic}.  For two configurations $X_1, X_2$ with pre-shapes $Z_1, Z_2$ respectively, the Riemannian shape distance is then defined as the minimum pre-shape distance between $Z_1$ and $Z_2\Gamma$, where the minimum is taken over $\Gamma \in SO(p).$  The following lemma from \cite{kendall1984shape} provides a representation of the optimal rotation in terms of the SVD of the pre-shape inner product $Z_1'Z_2 = U\Lambda V'$, $U, V \in SO(p)$.

\begin{lemma}[\cite{kendall1984shape}]
For pre-shapes $Z_1, Z_2$ with inner product SVD $Z_1'Z_2 = U\Lambda V', \Lambda = \mathrm{diag}(\lambda_1, \cdots, \lambda_p)$, the \textit{optimal rotation} $\Gamma \in SO(p)$ is
\begin{equation}\label{eqn:optimal_rotation}
    \hat \Gamma = \mathrm{argsup}~\mathrm{tr}\left(Z_1'Z_2\Gamma\right) = UV'.
\end{equation}
The value at the optimal rotation is
\begin{equation}\label{eqn:optimal_value}
    \sup_{\Gamma \in SO(p)} \mathrm{tr}\left(Z_1'Z_2\Gamma\right) = \sum_{i = 1}^p \lambda_i.
\end{equation}
\end{lemma}

\begin{proof} 
We first prove Equation \ref{eqn:optimal_value}, and then show that $\hat\Gamma$ in Equation \ref{eqn:optimal_rotation} attains this value.  Assume $\Gamma \in SO(p)$ has diagonal entries $\gamma_{ij}$.
\begin{align}
    \sup_{\Gamma \in SO(p)} \mathrm{tr}\left(Z_1'Z_2\Gamma\right) &= \sup_{\Gamma \in SO(p)} \mathrm{tr}\left(\Gamma\Lambda\right)\\
    &= \sup_{\Gamma \in SO(p)} \sum_{i = 1}^p \gamma_{ii}\lambda_i.
\end{align}

The set of diagonals of $\Gamma \in SO(p)$ is a convex set with extreme points $\{(\pm 1, \pm 1, \cdots, \pm 1)\}$ with $-1$ occurring an even number of times \citep{horn1954doubly}.  Consequently the maximum occurs for $\gamma_{ii} = 1$ for all $i = 1, \cdots, p$, giving the desired result.

From this result, plugging in $\hat\Gamma = UV'$ verifies Equation \ref{eqn:optimal_rotation}:
\begin{equation}
    \mathrm{tr}\left(Z_1'Z_2\hat\Gamma\right) = \mathrm{tr}\left(V\Lambda U'UV'\right) = \mathrm{tr}\left(\Lambda\right).
\end{equation}
\end{proof}

\begin{definition}
The \textbf{Riemannian shape distance} between two configurations $X_1$ and $X_2$ is equal to 
\begin{equation}
    \rho(X_1, X_2) = \arccos\left(\sum_{i = 1}^p \lambda_i\right),
\end{equation}
where $\lambda_i$ are the singular values of $Z_1'Z_2$ for the corresponding pre-shapes $Z_1, Z_2$.
\end{definition}

Intuitively, the shape distance of configurations $X_1, X_2$ is found by aligning the pre-shapes $Z_1, Z_2$ as closely as possible in the pre-shape sphere, and computing the arc length distance of the aligned pre-shapes.  In the language of manifold geometry, this is the geodesic distance between the fibers in the pre-shape space corresponding to the shapes $[X_1]$ and $[X_2]$.

\begin{theorem}
\label{th:Riemannian}
For configuration $X$ with unit disk polar representation $(r, \phi)$ and $X^*$ a configuration with shape equal to the B-midpoint triangle, the Riemannian shape distance between $X$ and $X^*$ is 
\begin{equation}
    \label{eqn:Riemannian}
    \rho(X,X^*) = \frac{1}{2}\arccos \{r \cos(\phi-\pi/3)\}.
\end{equation}
\end{theorem}

To establish the proof of Theorem \ref{th:Riemannian}, we first introduce two forms of Kendall's triangle coordinates to make use of previous results relating shape space representations and the Riemannian shape distance. To define the Kendall spherical coordinates, we first define rectangular Kendall coordinates for the case $p = 2$, which encompasses the general $p \geq 2$ case by mapping the plane containing a given triangle in $\mathbb{R}^p$ to $\mathbb{R}^2$.  Kendall's coordinates can be compactly expressed by considering the landmarks as points in the complex plane, $(z_1, z_2, z_3) \in \mathbb{C}^3$.  

\begin{definition}
The \textbf{Kendall's rectangular coordinates} for a triangular configuration $(z_1, z_2, z_3) \in \mathbb{C}^3$ are $(u_K, v_K)$ defined by $u_K + iv_K = \frac{z_{2}}{z_1}$.
\end{definition}

\begin{definition}
The \textbf{Kendall's spherical coordinates} $(\theta, \psi)$, $0 \leq \theta \leq \pi / 2, 0 \leq \psi \leq 2\pi$ are defined as a transformation of Kendall's rectangular coordinates
\begin{align}
    &\frac{1}{2}\sin\theta\sin\psi = \frac{u_K}{1 + r_K^2}\\
    &\frac{1}{2}\cos\theta = \frac{v_K}{1 + r_K^2},
\end{align}
where $r_K^2 = u_K^2 + v_K^2$.
\end{definition}

The vector of Cartesian coordinates in $\mathbb{R}^3$ for Kendall's spherical coordinates is
\begin{align}
    \ell &= \begin{pmatrix}
    \frac{1}{2}\sin\theta\cos\psi\\
    \frac{1}{2}\sin\theta\sin\psi\\
    \frac{1}{2}\cos\theta\end{pmatrix}.
\end{align}

Kendall's spherical coordinates are a representation of triangle shape space on the hemisphere with radius $\frac{1}{2}$ rather than the unit disk.  We focus on the disk representation in this work, as it provides for easier visualization of inferential results. The relation of the two representations is given in the following lemma.

\begin{lemma}
\label{lem:kendall_disk}
Kendall's spherical coordinates and the unit disk polar coordinates from Definition \ref{def:polar} are related by
\begin{align}
r &= \sin(\theta)\label{eqn:r_theta}\\
\phi &= \frac{2\pi}{3} - \psi\label{eqn:phi_psi}
\end{align}
\end{lemma}

Letting $\ell_1$ and $\ell_2$ be the Kendall coordinate vectors of the triangles $X_1, X_2$ respectively, we can write the Riemannian shape distance between $X_1$ and $X_2$ in terms of $\ell_1, \ell_2$ as
\begin{equation}
\cos \left(2\rho(X_1, X_2)\right) = 4\ell_1'\ell_2.
\end{equation}
Substituting the mappings \ref{eqn:r_theta} and \ref{eqn:phi_psi} yields the following proposition.
\begin{proposition}
\label{prop:r_rho}
For configurations $X_1, X_2$ with unit disk polar representation $(r_1, \phi_1), (r_2, \phi_2)$, the Riemannian distance is
\begin{equation}\label{eqn:r_rho}
    \rho(X_1, X_2) = \frac{1}{2}\arccos \left\{r_1r_2\cos(\phi_1 - \phi_2) + \sqrt{(1 - r_1^2)(1 - r_2^2)}\right\}.
\end{equation}
\end{proposition}

\begin{proof}[Proof of Theorem \ref{th:Riemannian}]
The result of Theorem \ref{th:Riemannian} is a direct consequence of Proposition \ref{prop:r_rho}, which is obtained by substituting $[X]=(r,\phi)$ and $[X^*]=(1,\pi/3)$ into Equation \ref{eqn:r_rho}.
\end{proof}

\subsubsection{Shape Space In-betweenness Index}
Note that the Riemannian distance $\rho(X,X^*)$ is between $0$ and $\pi/2$ with the minimum distance $0$ occurring uniquely when the configuration has shape equal to the B-midpoint triangle. Since this distance is intrinsic to shape space, it provides a natural and theoretically motivated means of defining a shape space similarity measure.
When quantifying the strength of a relationship via a similarity metric, it is desirable for practical interpretation to have a measurement with 1 for the strongest positive relationship and -1 for the strongest negative relationship. For example, both the cosine in-betweenness in Equation \ref{eq:gamma} and the Pearson's correlation coefficient satisfy this requirement. For this purpose, we transform the Riemannian distance in Equation \ref{eqn:Riemannian} using the decreasing function $\cos(2\rho(X,X^*))$.  This transformation of the Riemannian shape distance also has compact alternative representations in terms of the Euclidean inner product in shape space, and the triangle side lengths, which are detailed below.

\begin{definition}
\label{eqn:IBI_def}
The \textbf{shape space in-betweenness index (IBI)} $\tau$ measuring the in-betweenness of group B with respect to groups A and C from an observed triangle $X$:
\begin{equation}
\tau = \cos(2\rho(X, X^*)),
\end{equation}
where $X^*$ is a configuration of the B-midpoint triangle.
\end{definition}

\noindent By definition, the maximum $\tau$ is uniquely attained when $(r, \phi) = (1, \pi / 3)$.  The minimum $\tau = -1$ is uniquely attained at $(1, -\pi / 2 + \pi / 3)$, which corresponds to $X_A = X_C \neq X_B$.  All triangles with $\tau = 0$ lie on the line defined by $\phi = \pi/ 3 \pm \pi / 2$ (or equivalently $u = -\sqrt 3 v$), which includes the equilateral triangle at $(0, 0)$.  

We note that, as a consequence of Proposition \ref{prop:r_rho}, when one of the two triangles is degenerate, i.e., the radius is $1$, the Riemannian distance between the two triangles equals the Euclidean inner product of their unit disk shape space representations.  
\begin{corollary}\label{cor:rho}
When at least one of the triangles determined by $X_1, X_2$ is degenerate, say $r_2 = 1$, the Riemannian distance reduces to
\begin{equation}
    \rho(X_1, X_2) = \frac{1}{2}\arccos \langle X_1, X_2\rangle_{\mathcal{U}},
\end{equation}

\noindent where $\langle X_1, X_2\rangle_{\mathcal{U}} = r_1r_2\cos(\phi_1 - \phi_2)$ is the Euclidean inner product of the unit disk shape space representations of $X_1$ and $X_2$.
\end{corollary}

\noindent $\tau$ is initially motivated from the pre-shape space Riemannian distance, as the distance is intrinsic in the pre-shape Riemannian manifold. It has several alternative forms that are of interest. Firstly, rewriting Definition \ref{eqn:IBI_def} in terms of rectangular shape coordinates in Definition \ref{def:polar} gives $\tau = \frac{1}{2}u + \frac{\sqrt{3}}{2}v$.  Secondly, substituting the expressions for $u, v$ in terms of $a^2, b^2$ yields the simplification $\tau = 3b^2 - 1$, which indicates that the hybridity measure is a transformation of the scaled side length $b^2$. These connections are summarized in the follow theorem.  

\begin{theorem}
\label{prop:tau_pdf}
For a triangle with configuration $X$, scaled side lengths $a^2, b^2, c^2$, coordinates $(r\cos\phi, r\sin\phi)$, and Riemannian distance to the $B$-midpoint triangle $\rho$, the shape space IBI $\tau = \cos(2\rho(X, X^*))$ has the equivalent forms:
\begin{enumerate}
    \item $\tau = r\cos(\pi / 3 - \phi)$
    \item $\tau = \frac{1}{2}u+\frac{\sqrt{3}}{2}v$
    \item $\tau = 3b^2 - 1$
\end{enumerate}
\end{theorem}

\begin{proof}

\begin{enumerate}
    \item This is from Theorem \ref{th:Riemannian}
    \item \begin{align*}
        \tau &= r\cos\left(\phi - \frac{\pi}{3}\right)\\
        &= r\cos\phi\cos\frac{\pi}{3} + r\sin\phi\sin\frac{\pi}{3}\\
        &= \frac{1}{2}u + \frac{\sqrt{3}}{2}v
    \end{align*}
    \item Computing $b^2$ from Equation \ref{eq:uv_ab} gives $b^2 = \frac{1}{3}(\frac{1}{2}u + \frac{\sqrt{3}}{2}v) + \frac{1}{3}$, which implies $\tau = 3b^2 - 1$.
\end{enumerate}
\end{proof}

\begin{corollary} When the configuration $X$ has landmark distribution $X_i \sim \mathcal{N}(0, \sigma^2I_p)$, $i = A, B, C$, the density of $\tau$ is $f_{\tau}(t) = \frac{\Gamma(\frac{p + 1}{2})}{\sqrt{\pi}\Gamma(\frac{p}{2})} (1 - t^2)^{(p - 2)/2}$ for $-1 \leq t \leq 1$.
\end{corollary}

\noindent \textbf{Proof} From Theorem \ref{prop:tau_pdf}, the null distribution of $\tau$ is straightforward to compute as a transformation of $3/2 b^2 \sim \text{Beta}(p/2, p/2)$ \citep{edelman2015random}.

\begin{figure}
    \centering
    \includegraphics[width=0.9\textwidth]{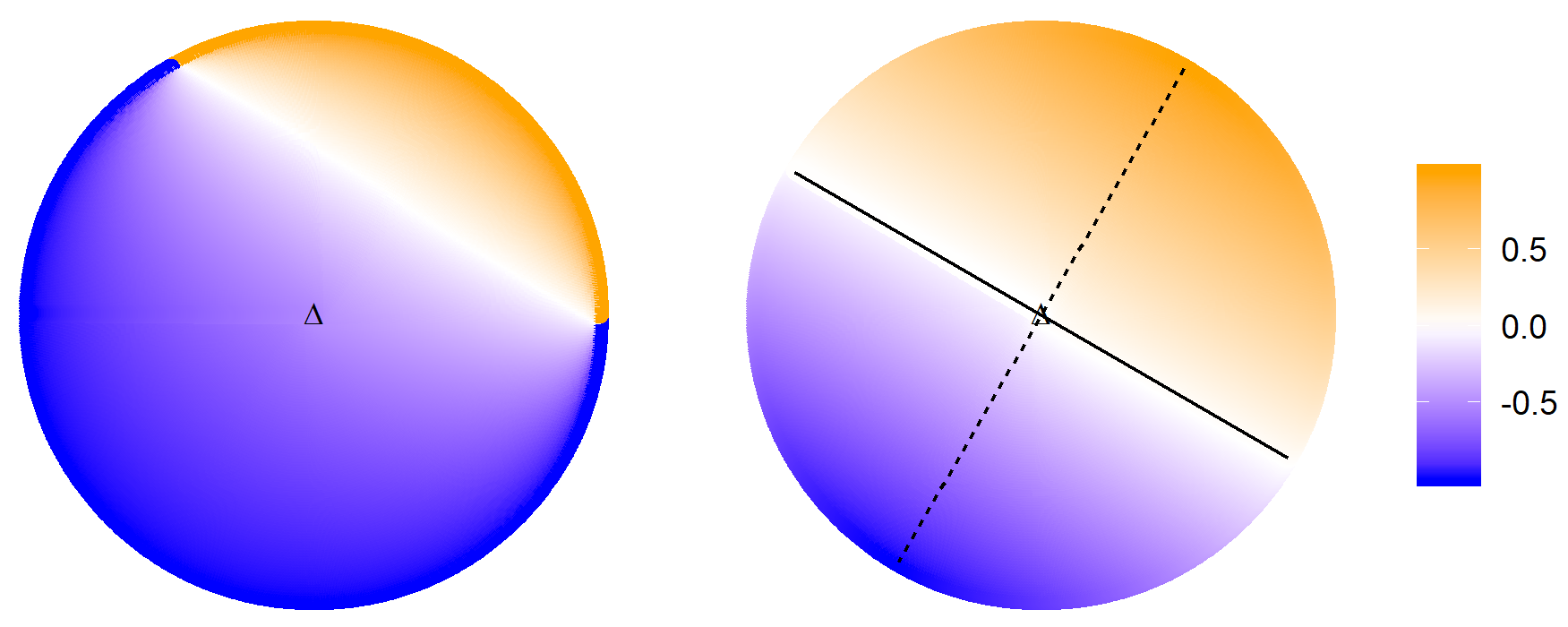}
    \caption[Shape space plot of $\gamma$ and $\tau$ in-betweenness indices]{(\textit{left}) Value of cosine IBI over triangle shape space.  This measure has discontinuities on the boundary at $\phi = 0$ and $\phi = 2\pi / 3$, where $B$ switches from in-between $A$ and $C$ to outside the $AC$ segment. (\textit{right}) Value of $\tau$ IBI over triangle shape space.  This measure is continuous in the entire space.  The maximum $\tau = 1$ occurs at $(r, \phi) = (1, \pi / 3)$, which corresponds to the $B$-midpoint triangle.}
    \label{fig:IBI_values}
\end{figure}

\textbf{Remark} A comparison of the values of $\gamma$ and $\tau$ the unit disk is shown in Figure \ref{fig:IBI_values}.  Because of the discontinuities and insensitivity to different forms of in-betweenness, the $\tau$ in-betweenness measure should generally be preferred, although $\gamma$ may be of use when the positioning of $B$ along the $AC$ edge is not important.

\subsubsection{Offset-Normal Distributions}
The Riemannian distance result can also be used to derive the distribution for isotopic case with non-coincident landmark centroids. In the isotropic case with non-coincident landmark centroids, $X \sim \mathcal{N}(\mu, \sigma^2 I_3, I_p)$, the \textit{off-set normal shape density} in terms of the Riemannian shape distance $\rho(X, \mu)$ is 
\begin{equation}
    \left\{1 + \kappa\left[1 + \cos(2\rho(X, \mu))\right]\right\}\exp\left\{-\kappa\left[1 - \cos(2\rho(X, \mu))\right]\right\},
\end{equation}
where $\kappa = S^2(\mu) / (4\sigma^2)$, for population centroid size $S(\mu)$ \citep{mardia1989statistical}. The density in terms of polar shape coordinates can be derived via variable transformation and using the results from Proposition \ref{prop:r_rho}.  In the context of analyzing three-group data, this distribution is only applicable when the sample is balanced across the groups. In the unbalanced case, the isotropic assumption is violated for the group centroids, thus limiting the practical use of the isotropic offset normal distribution.

Some results for distributions with general covariances have been derived, but known expressions of the shape distribution are complicated, involving finite sums of generalized Laguerre polynomials.  A detailed discussion of the offset normal distribution is given in \cite{dryden2016statistical}, which includes the density function for triangle shape when $p = 2$ and the configuration distribution is a complex normal with general covariance. In practice, resampling methods are often adequate for most inference purposes. In the following, we propose a bootstrap procedure to make inference on shape space parameters such as $\tau$ and shape space location.

\subsection{Stratified Bootstrap Procedure for Shape Space Inference}

A bootstrap approach for creating $\tau$ confidence intervals is suitable when the covariance structure in each stratum can be assumed to be exchangeable.  Although we focus here on inference for $\tau$, the same algorithm can be extended to provide inference for the cosine in-betweenness $\gamma$, shape space coordinates, and side lengths.

IBI quantification is of greatest interest when it is suspected that the three group centroids are not coincident; when the centroids are in fact nearly coincident (relative to the variance in the data), the bootstrap distribution closely approximates the null distribution for the IBI statistic, rather than concentrating around $\tau = 0$.

\begin{algorithm}
\caption{Stratified Bootstrap for Shape Analysis}\label{alg:ibt_bootstrap}
\begin{algorithmic}[1]
\State Input: $X_1, X_2, X_3$, $K$
\For{$k$ in $1:K$}
\State Create stratified bootstrap replicate $X_1^{(k)}, X_2^{(k)}, X_3^{(k)}$.
\State Compute bootstrap configuration $\bar X_1^{(k)}, \bar X_2^{(k)}, \bar X_3^{(k)}$.
\State Compute normalized side lengths $a_k^2, b_k^2, c_k^2$.
\State Compute bootstrap IBI statistic $\tau^{(k)} = 3 b_k^2 - 1$.
\EndFor
\State Compute the bootstrap IBI $100(1 - \alpha)\%$ confidence interval as the $\alpha/2$ and $1 - \alpha/2$ percentiles of $\{\tau^{(k)} | k = 1, \dots, K\}.$
\end{algorithmic}
\end{algorithm}

Confidence regions for the shape space coordinates can be computed as a byproduct of the bootstrap procedure by recording the bootstrap sample quantities $(u^{(k)}, v^{(k)})$.  Confidence regions from these bootstrap samples can then be computed using a data depth metric, such as Tukey data depth \citep{di2004multivariate}.  The shape space confidence region can provide greater insight into the likely relationship of the three subpopulations through inspection of the extreme triangles.

\textbf{Remark} Similar to principal components analysis, the question of whether to standardize the features to unit variance before conducting shape space inference should be considered carefully with respect to the scientific meaning of the features and the inference goal.  Unless the observed features have equal sample variance, standardization will scale each feature different, with the potential to substantially alter the estimated shape and confidence region.  Throughout this work we consider standardized features, but note that not standardizing may be more appropriate in some settings, particularly when all of the features are of a similar type and measured on the same scale.

\section{Simulations \& Applications}

\subsection{Simulations}
\label{triangles:simulations}
To evaluate the proposed bootstrap method for IBI quantification, we simulate data with features generated from: i) a standard normal distribution with (potentially) different group centroids; ii) a standard normal distribution with varying sample sizes across groups. 

The size and coverage of bootstrap 95\% confidence regions for shape space location and confidence intervals for $\tau$ were assessed using $1000$ replications of balanced data generated from an isotropic normal distribution with mean configuration specified by $r = 0.5, \phi = \pi / 3$, for $n = 90, 300$ and $\sigma^2 = 0.1, 1, 5$.  The simulation results (Table \ref{tab:coverage_simulation_results}) show approximately correct coverage for the $\tau$ confidence intervals across all settings; the confidence regions for shape space location perform slightly worse, with coverage around 93\% for most settings.  The results also show substantial contraction of the confidence regions and intervals as $n$ increases and $\sigma^2$ decreases.

\begin{table}
\caption{Simulation results for coverage of stratified bootstrap. Simulation results for data sample iid from a normal distribution with mean configuration specified by $r = 0.5, \phi = \pi / 3, p = 2$.  Confidence intervals for $\tau$ and confidence regions for $(u, v)$ were calculated with the stratified bootstrap with 1000 simulation replications and $2000$ bootstrap replications per simulation.}
\label{tab:coverage_simulation_results}
\centering
\begin{tabular}{cccccc}
\toprule
    $n$ & $\sigma^2$ & CI Cover. & CI Length & CR Cover. & CR Area\\
    \midrule
        90 & 5 & 0.952 & 1.233 & 0.978 & 1.951\\
        300 & 5 & 0.956 & 0.795 & 0.949 & 0.839\\
        90 & 1 & 0.938 & 0.662 & 0.939 & 0.574\\
        300 & 1 & 0.953 & 0.381 & 0.933 & 0.193\\
        90 & 0.1 & 0.938 & 0.22 & 0.927 & 0.064\\
        300 & 0.1 & 0.952 & 0.123 & 0.932 & 0.02\\
    \bottomrule
\end{tabular}
\end{table}

\subsection{Shape Analysis of Iris Data}

\begin{figure}
    \centering
    \includegraphics[width=\textwidth]{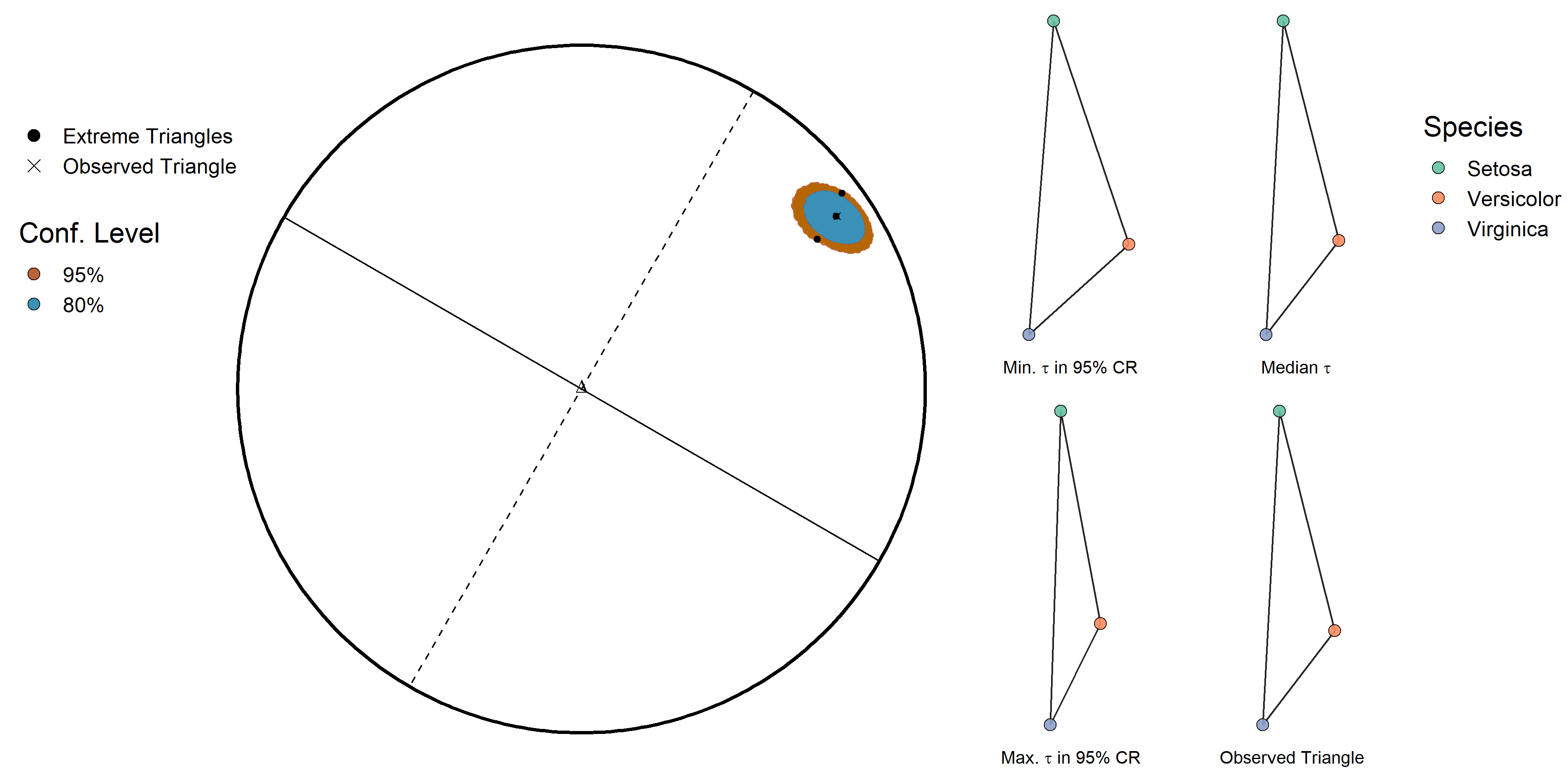}
    \caption[Shape space confidence regions for iris data, all features]{80\% and 95\% bootstrap CRs for iris data, for \textit{versicolor} as group $B$, using all four features. There is some indication that the mean \textit{versicolor} features lie approximately between the mean features for \textit{setosa} and \textit{virginica}, corroborating previous evidence that the mean \textit{versicolor} features are approximately in-between two other species.}
    \label{fig:iris_extreme}
\end{figure}

For a simple illustration of the statistical shape analysis and the IBI statistics, we consider the classic iris data set \citep{fisher1936use, anderson1936species}.  This data set provides a convenient example of in-betweenness analysis, as it consists of three iris species, one of which (\textit{versicolor}) is believed to be a genetic hybrid of the others (\textit{setosa} and \textit{virginica}).  To quantify this relationship as manifested in physical characteristics, we calculate 95\% bootstrap confidence region for shape, and confidence intervals for $\tau$ and $\gamma$, for the four standardized features (sepal width and length, and petal width and length), using $10000$ bootstrap replications.  The 80\% and 95\% bootstrap confidence region and the extreme triangles from the 95\% with the maximum and minimum $\tau$ are shown in Figure \ref{fig:iris_extreme}.  While the $\tau$ measure provides one indication of position in shape space, it can be difficult to interpret directly, thus we recommend also examining the confidence region boundary shapes and median shape estimate in order to better understand the range of likely shapes.

\begin{table}
\caption{Confidence intervals for in-betweenness indices for iris data. Observed $\tau$ and $95\%$ CI for different subsets of features from the iris data set, measuring \textit{versicolor} as a hybrid of \textit{setosa} and \textit{virginica}. Features include sepal length (SL), sepal width (SW), petal length (PL), and petal width (PW). There is strong evidence that the mean \textit{versicolor} features lie between the centroids for \textit{setosa} and \textit{virginica}, with $\tau = 0.909$ over all four features.}
  
\label{tab:iris_IBI}    
\centering
    \begin{tabular}{lcc}
    \toprule
        Features & Obs. $\tau$ (95\% CI) & Obs. $\gamma$ (95\% CI)\\
        \midrule
         SL, SW &  0.817 (0.732, 0.872) & 0.103 (-0.182, 0.448)\\
         SL, PL &  0.922 (0.885, 0.949) & 0.979 (0.936, 0.9996)\\
         SL, PW & 0.974 (0.936, 0.990) & 0.999 (0.997, 0.999)\\
         All features & 0.909 (0.879, 0.931) & 0.624 (0.444, 0.795)\\
   \bottomrule
    \end{tabular}
\end{table}

\begin{figure}
    \centering
    \includegraphics[width=\textwidth]{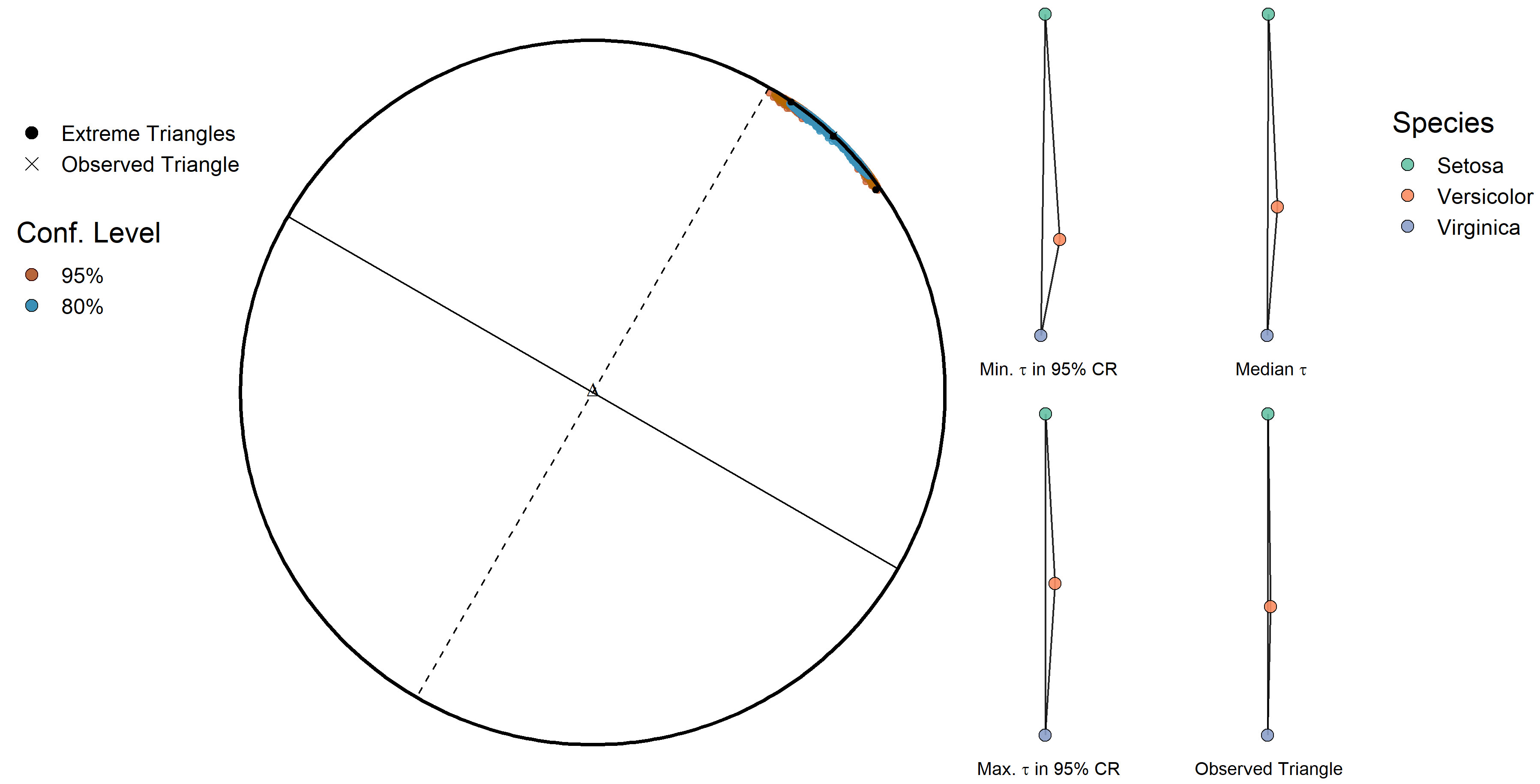}
    \caption[Shape space confidence regions for iris data, selected features with maximum collinearity]{80\% and 95\% bootstrap CRs for the sepal length and petal width features, with \textit{versicolor} as group $B$. The observed triangle is very close to degenerate, resulting in the confidence regions being distributed as narrow bands along the shape space boundary.  The confidence regions are strongly indicative of near collinearity of centroids.}
    \label{fig:iris_extreme_max_features}
\end{figure}

From our simulations and real-world analyses, we have observed that the bootstrap confidence regions are elliptically shaped when the observed triangle is not near the shape space boundary and the variance in the data is not too large relative to the observed centroids.  When the observed triangle is instead close to the shape space boundary (i.e.~approximately degenerate) and the variance is not too large, the estimated confidence regions tend to be distributed as a narrow band along the boundary.  As an example of is, Figure \ref{fig:iris_extreme_max_features} shows the confidence region and extreme triangles for iris features with maximum $\tau$, sepal length and petal width.  

\subsection{PAM50 Breast Cancer Data}

In our second application example, we investigate a data set from an analysis of hormone receptor-positive breast cancer subtypes and risk of relapse \citep{prat2017pam50}.  The subtypes considered include Luminal A (\textit{LumA}), Luminal B (\textit{LumB}), and Basal-like. This study conducted a meta-analysis of the relation of a genomic-based chemoendocrine score (CES) with risk of relapse (ROR) across 6007 tumors, finding that CES estimates of chemoendocrine sensitivity beyond what is indicated by the intrinsic cancer subtype and clinical covariates.  A primary result of this study is evidence that sensitivity to endocrine therapy and chemotherapy is linked to the biological differences in Basal-like versus Luminal A subtypes.  Given the strong association with chemosensitivity and risk of relapse, there is interest in better understanding the relative relationships of these three subtypes \citep{prat2017pam50}. Toward this end, we generate the shape space stratified bootstrap confidence regions to describe the relative relationships of the three subtypes with respect to the CES and ROR measures. 

\begin{figure}
    \centering
    \includegraphics[scale=0.6]{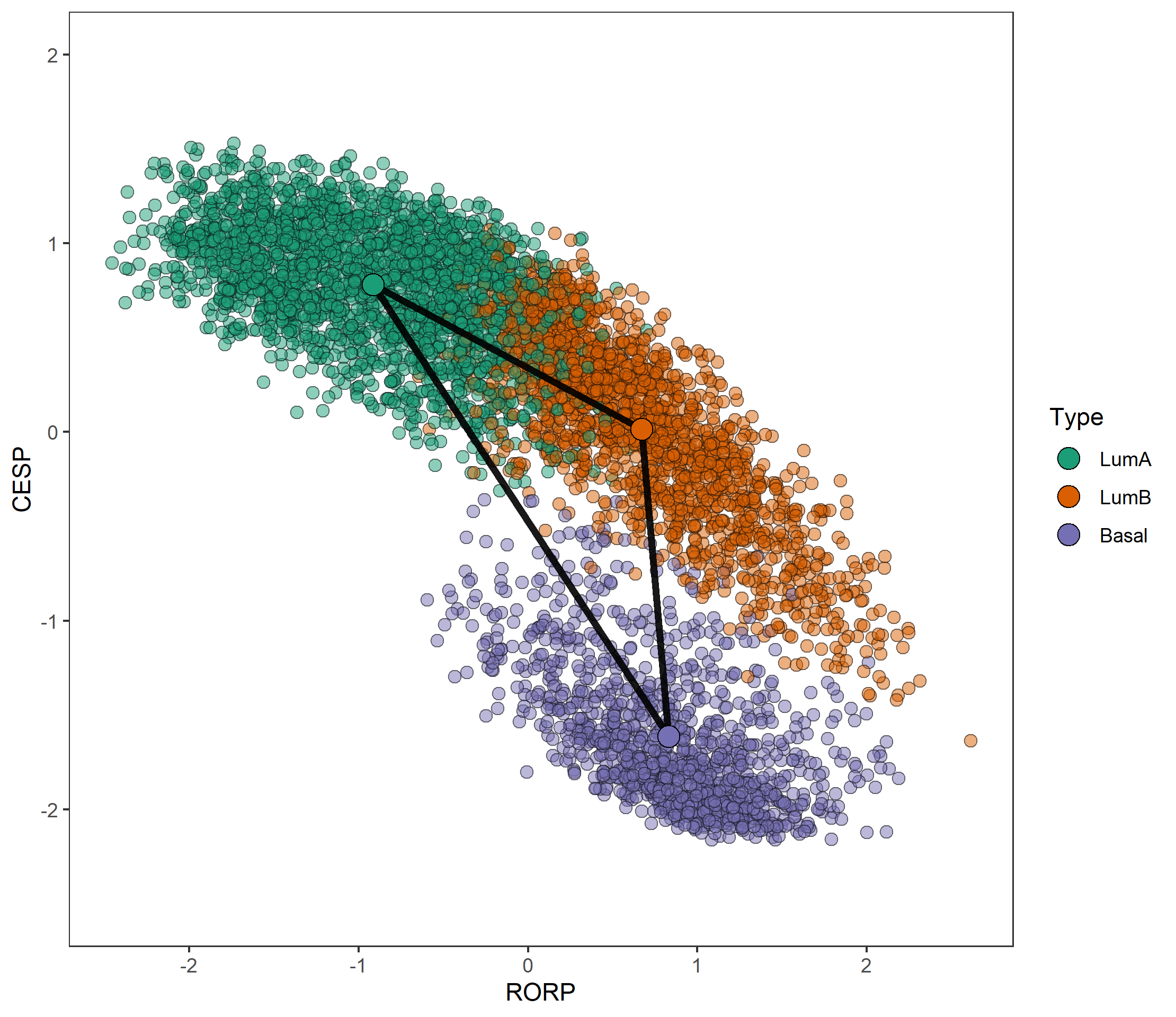}
    \caption[Plot of observations from PAM50 breast cancer meta-analysis]{Data from the meta-analysis of breast cancer subtypes identified by the PAM50 genetic indicator.}
    \label{fig:pam50_obs}
\end{figure}

A plot of the PAM50 data set from \cite{prat2017pam50} is given in Figure \ref{fig:pam50_obs}.  The joint centroids across the CES and ROR features are clearly non-coincident for the three groups, with each subtype forming a distinct cluster. Overall lower risk of relapse is apparent in the \textit{LumA} group, with approximately similar distribution of ROR for \textit{LumB} and \textit{Basal}. Chemosensitivity shows an approximate linear relationship with ROR across the \textit{LumA} and \textit{LumB} groups, with the \textit{LumA} group showing a higher mean CES than \textit{LumB}, but is distinctly lower for the \textit{Basal} group.

\begin{figure}
    \centering
    \includegraphics[width=\textwidth]{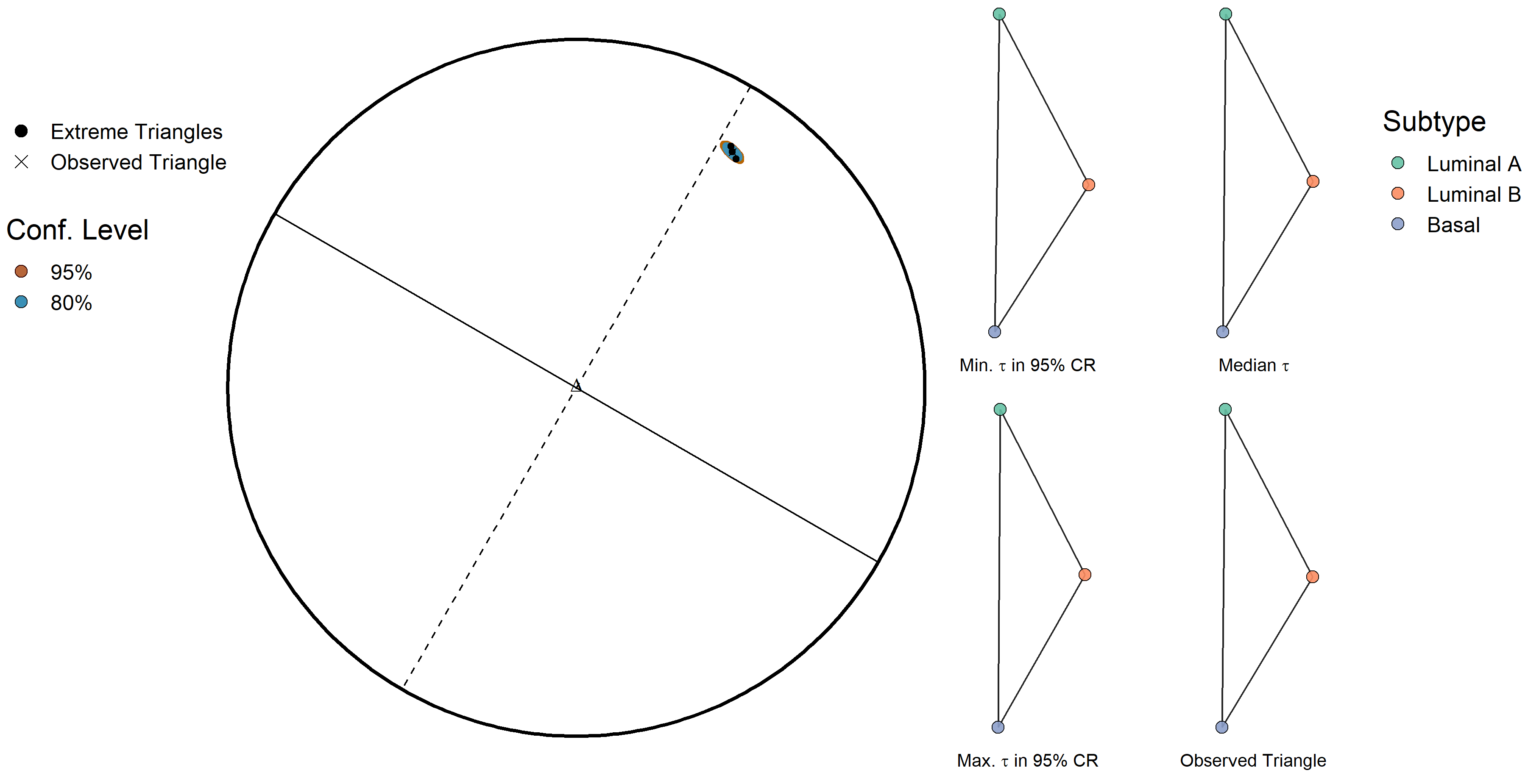}
    \caption[Shape space inference results for PAM50 data]{The observed triangle, and the median, and extreme triangles in the 95\% bootstrap CR for the PAM50 data set show little variation, indicating strong evidence that the three group centroids are approximately collinear, with the \textit{LumB} mean close to the \textit{LumA} mean, and between the centroids for \textit{LumA} and \textit{Basal} groups.}
    \label{fig:pam50_extreme}
\end{figure}

We construct the shape space bootstrap confidence region for the PAM50 data set, taking \textit{LumA, LumB,} and \textit{Basal} as the $A, B,$ and $C$ groups respectively, using $5000$ bootstrap permutations.  The 95\% bootstrap confidence region and corresponding extreme triangles are shown in Figure \ref{fig:pam50_extreme}.  The bootstrap median and 95\% confidence interval for $\tau$ is $0.810$  $(0.800, 0.819)$, with an observed $\tau = 0.810$; the $\gamma$ median and 95\% confidence interval are $0.522$ $(0.502, 0.541)$, with an observed $\gamma = 0.522$.  The concentration of the confidence region and similarity of the extreme triangles in this region (Figure \ref{fig:pam50_extreme}) provide strong evidence that the observed shape is very close to the true mean shape.  Compared to the iris example above, we see the confidence region from the PAM50 results is much more concentrated due to the larger sample size. 

\subsection{CCK/PV Cell Data}

\begin{figure}
    \centering
    \includegraphics{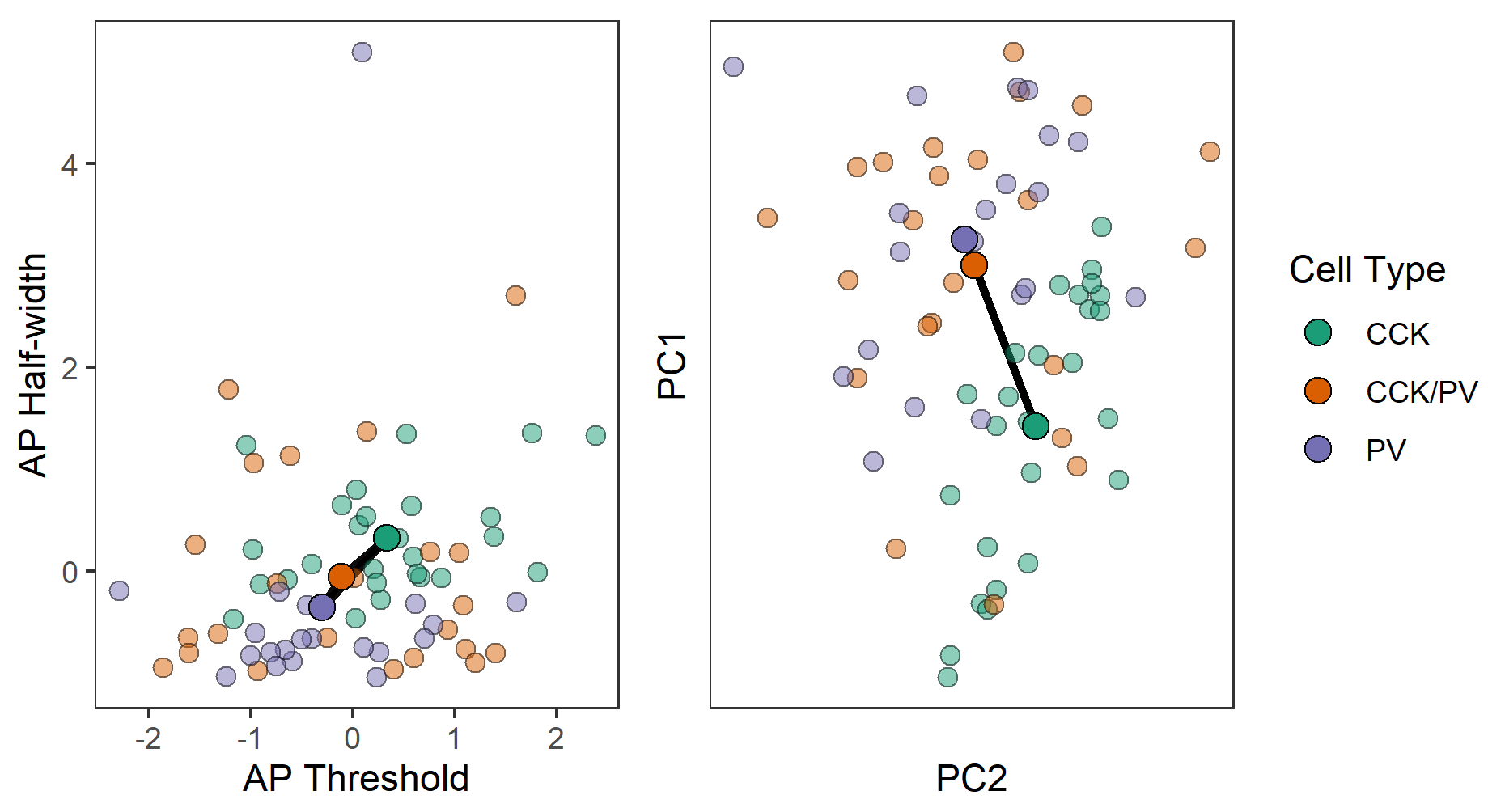}
    \caption[Plots of observations from the CCK/PV data set]{(\textit{left}) Plot of observations from the CCK/PV data set for AP threshold and AP half-width features. (\textit{right}) Plot of observations for the first two principal components calculated from all 12 measured features.}
    \label{fig:cell_obs}
\end{figure}

The development of the IBI methodology here is motivated by a study of mouse hippocampal CCK and PV neurons.  Novel cells co-expressing CCK and PV are have been discovered in mice, but not rats. It is of interest to evaluate the ``in-betweenness'' of the electrophysiological characteristics of CCK/PV cells with respect the individually expressing CCK and PV cells.  For this study, the 12 measured electrophysiological features are action potential (AP) frequency, AP amplitude, AP  threshold, AP adaptation index, AP risetime, AP half width, AP falltime, after hyperpolarization potential (AHP), AHP time, resting membrane potential (RMP), input resistance and hyperpolarization current (-100pA) induced inward rectification ``sag.''  Sample sizes of measured interneurons from each group are $n = 23$ CCK+/PV+, $n = 26$ CCK+/PV-, and $n = 20$ CCK-/PV+ (n=19). Figure \ref{fig:cell_obs} shows the CCK/PV observations for AP half width against AP threshold, and for the first two principal components calculated from all features.  We note that the cell type groups are not clearly clustered, and that, for most of the measured features, the variance of the observations is large relative to the distance between group centroids.

We consider this data from the shape space perspective, and compare $\tau$ to the cosine similarity. To assess the strong null hypothesis that all moments across the three cell groups are equal, we conduct a permutation hypothesis test by shuffling group labels to generate 5000 permuted data sets and calculate the cosine similarity and IBI statistic for each permutation.  The resulting $P$-values are $P_{\gamma} = 6 \times 10^{-4}, P_{\tau} = 0.0072$, indicating strong evidence that the cell group centroids are not coincident.

\begin{figure}
    \centering
    \includegraphics[width=\textwidth]{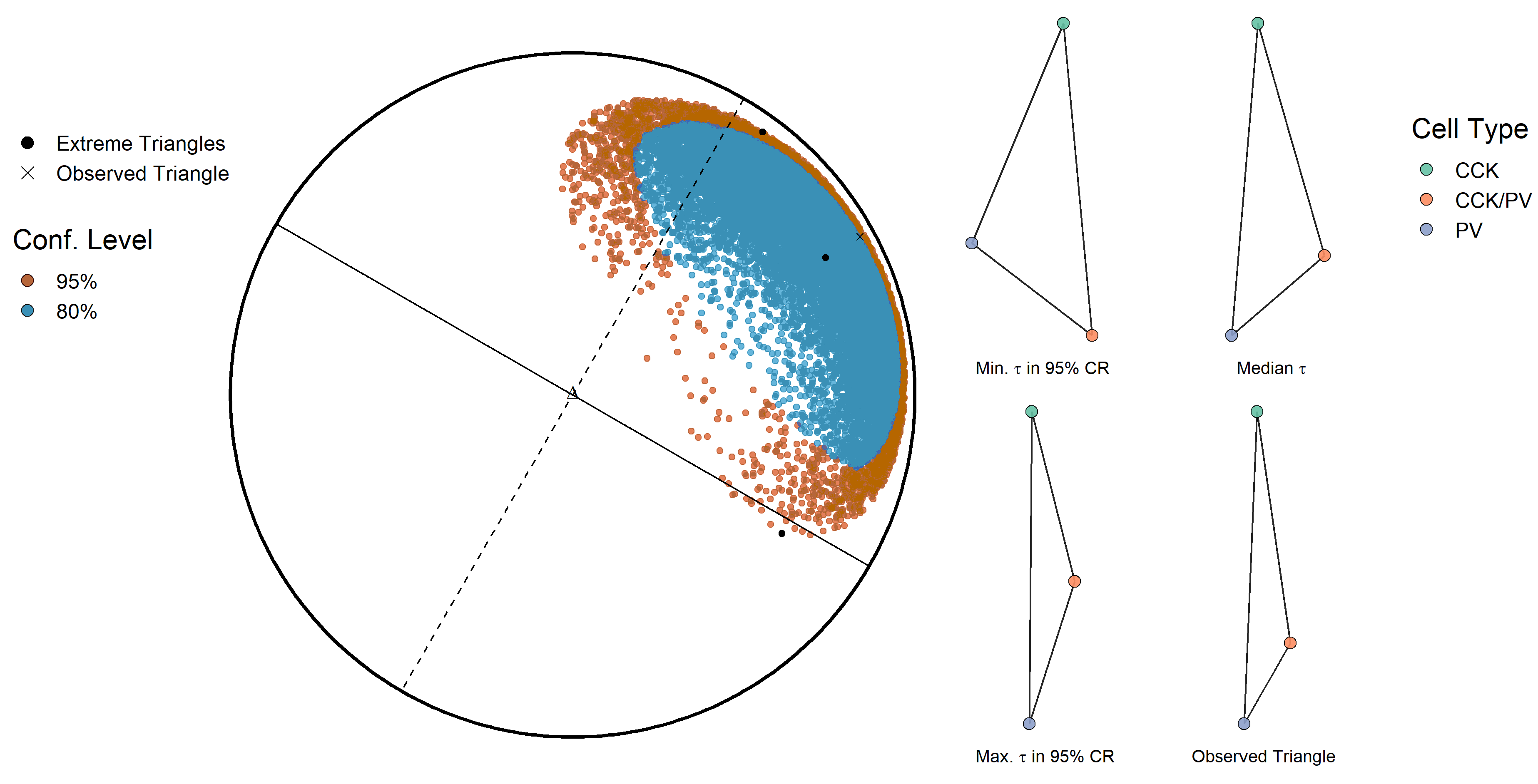}
    \caption[Shape space inference results for CCK/PV data]{Observed shape, and median and extreme shapes from the 95\% bootstrap CR for CCK/PV cell data.  Due to the variance in the observed features, and small sample sizes across groups, there is significant variation in the likely shapes.  There is some indication that the CP mean is approximately between the C and P centroids (as in the maximum $\tau$ triangle), but the minimum $\tau$ triangle is not suggestive of collinearity of centroids.}
    \label{fig:cell_extreme}
\end{figure}

The shape space stratified bootstrap procedure (Algorithm \ref{alg:ibt_bootstrap}) provides a description of the likely triangles formed by the cell group centroids.  The shape space confidence regions (Figure \ref{fig:cell_extreme}) show a wide range of possible shapes, resulting from the large variance in the data and relatively small sample sizes.  Examining the extremal triangles in the 95\% CR, we see that there is wide variation in the possible mean shapes.  From the median triangle and 95\% CR triangle with maximum $\tau$, there is some indication that the CP mean is approximately between the C and P centroids, however the minimum $\tau$ triangle is not suggestive of collinearity of centroids.  Thus, although the observed triangle is approximately collinear with the CP mean between the C and P centroids, the data do not provide sufficient evidence to conclude that the CP group mean lies approximately between the other group centroids.  

\begin{table}
    \caption[Confidence intervals for in-betweenness indices for CCK/PV data]{95\% confidence intervals for $\gamma$ and $\tau$ in-betweenness indices for the CCK/PV data, computed with 20000 bootstrap replications. The $\tau$ confidence interval does not contain 0, allowing us to conclude that the CCK/PV centroids do not have an equilateral relationship.  However, the width of this interval makes it difficult to say more regarding the relationship of the group centroids.}    \centering
    \begin{tabular}{cccc}
    \toprule
        IBI Type & Median (95\% CI) & Observed IBI\\\midrule
        $\gamma$ & 0.461 (-0.562, 0.837) & 0.820\\
        $\tau$ & 0.722 (0.213, 0.915) & 0.789\\
        \bottomrule
    \end{tabular}

    \label{tab:cell_results}
\end{table}

\section{Discussion}

Although the theory of statistical shape analysis has been thoroughly developed in the context of observed samples of shapes, relatively little attention has been given to the study of configurations of summary statistics arising from multiple observed subpopulations.  The proposed $\tau$ IBI provides a one-dimensional measure of shape space location such that the $B$-midpoint triangle maximizes $\tau$, thus $\tau$ values close to 1 indicate triangles for which the $B$ subpopulation mean is approximately equal to the midpoint of centroids for subpopulations $A$ and $C$.  Similar in spirit to correlation measures, the $\tau$ IBI provides a point of reference in evaluating the in-betweenness exhibited by a particular sample, and may be useful as a point of comparison across studies or samples.  However, since interpretation of $|\tau| << 1$ may be difficult, it is useful to also consider shape space confidence regions to describe the range of likely triangles.  These shape space confidence regions, as constructed by the stratified bootstrap procedure used here, provide greater insight into the possible shapes formed by the subpopulation centroids.  Specifically, through consideration of the extremal and median triangles in the confidence region, one may investigate the relative orderings and range of likely relationships of the subpopulation centroids.  In ideal situations, with small variation in confidence region triangles, it may be possible for researchers to conclude that the subpopulation centroids exhibit a particular relationship of interest.

The shape space framework offers many advantages when the scientific question of interest concerns the relative positioning.  The inference methods developed here can be applied to an arbitrary number of features, and allow for convenient visualization of the uncertainty in relative mean positions regardless of the ambient dimension of the feature space.  As the above simulation results show, the performance of the shape space methods are robust to increasing dimension, and in fact the permutation test for coincident centroids using $\tau$ or cosine IBI show increased power as dimension increases, as a result of the null distribution concentrating around the shape space origin.  

A potential drawback of shape space approaches is the need for bootstrap or other randomization methods for the construction of confidence regions, due to the complexity of the shape space distributions in the non-null cases.  However, for sample sizes common in many biological and medical studies, the required computation is generally tractable.  As the underlying computations are routine linear algebra operations, greater computational efficiency can be achieved through the use of specialized hardware and linear algebra software packages.

There are many possible extensions and improvements on the methods developed here.  While the present work has focused solely on triangle shape space methods for the analysis of three subgroups, the ideas may be extended to study the relative relationships of more than three groups.  Although the coverage of the stratified bootstrap shows good performance in the simulation settings considered here, alternative bootstrap procedures may be considered to reduce bias in the bootstrap estimates, e.g. a double bootstrap or bias-corrected bootstrap.

\bibliographystyle{rss}
\bibliography{references}
\end{document}